\documentclass[twocolumn,pra,superscriptaddress,noshowpacs,epsf]{revtex4}

\usepackage{mathrsfs}

\usepackage{graphicx}

\begin{document}
\title{Topological quantum phase transition in the extended Kitaev spin model}
\date{\today}%
\author{Xiao-Feng Shi}
\affiliation{Department of Physics and Surface Physics Laboratory
(National Key Laboratory), Fudan University, Shanghai 200433, China}
 \affiliation{Advanced Science Institute, The Institute of Physical and Chemical Research (RIKEN), Wako-shi 351-0198, Japan}
\affiliation{CREST, Japan Science and Technology Agency, Kawaguchi,
Saitama 332-0012, Japan}
\author{Yue Yu}
\affiliation{Institute of Theoretical Physics, Chinese Academy of
Sciences, P.O. Box 2735, Beijing 100190, China}

\author{J. Q. You}

 \affiliation{Department of Physics and Surface Physics Laboratory
(National Key Laboratory), Fudan University, Shanghai 200433, China}
 \affiliation{Advanced Science Institute, The Institute of Physical and Chemical Research (RIKEN), Wako-shi 351-0198, Japan}
\affiliation{CREST, Japan Science and Technology Agency, Kawaguchi,
Saitama 332-0012, Japan}
\author{Franco Nori}

 \affiliation{Advanced Science Institute, The Institute of Physical and Chemical Research (RIKEN), Wako-shi 351-0198, Japan}
\affiliation{CREST, Japan Science and Technology Agency, Kawaguchi,
Saitama 332-0012, Japan}
 \affiliation{Center for Theoretical Physics, Physics Department, Center for the Study of Complex Systems,University of Michigan, Ann Arbor, MI 48109-1040, USA
}

\begin{abstract}
We study the quantum phase transition between Abelian and
non-Abelian phases in an extended Kitaev spin model on the honeycomb
lattice, where the periodic boundary condition is applied by placing
the lattice on a torus. Our analytical results show that this spin
model exhibits a continuous quantum phase transition. Also, we
reveal the relationship between bipartite entanglement and the
ground-state energy. Our approach directly shows that both the
entanglement and the ground-state energy can be used to characterize
the topological quantum phase transition in the extended Kitaev spin
model.
\end{abstract}

\maketitle

\section{introduction}
Quantum phase transitions, which occur when a driving parameter in
the Hamiltonian of the system changes across a critical point, play
a central role in condensed matter physics~\cite{continuous phase
transition,quantum phase transition}. While most quantum phase
transitions can be characterized by symmetry breaking, there is also
an exception that can only be witnessed by topological order (see,
e.g., \cite{xgwen02,xgwen01}). Signatures of topological order in
many-body quantum systems can characterize a topological quantum
phase transition and include, e.g., the existence of excitations
obeying fractional statistics (see, e.g.,~\cite{quasi}),
ground-state degeneracy related to the topology of the system
(instead of the symmetry) (see, e.g.,~\cite{xgwen01,deg01}), and
topological entanglement entropy~\cite{topoentropy01,topoentropy02}.
In particular, the spectral Chern number~\cite{Kitaev} serves as a
topological number for characterizing a two-dimensional (2D) system
of noninteracting (or weakly interacting) fermions with an energy
gap. Without closing the gap, energy spectra with different Chern
numbers cannot be deformed into each other~\cite{DHLee}. This is
because a topological quantum phase transition occurs when changing
the Chern number.

Recently, it was shown~\cite{XYFeng,HDChen} that the topological
quantum phase transition in the Kitaev spin model can be
characterized by nonlocal-string order parameters. In an
appropriate dual representation, this order parameter can become
local and the basic concept of Landau theory of continuous phase
transition is also applicable~\cite{XYFeng}.

\begin{figure}
\includegraphics[width=3.3in,bbllx=0,bblly=368,bburx=589,bbury=590]{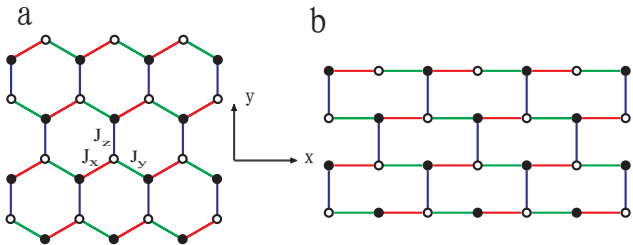}
 \caption{(Color online) (a) Honeycomb lattice constructed by two
 triangular lattices that are connected to each other by three types of bonds $J_x, J_y$ and
 $J_z$. (b) The brick-wall lattice, which is deformed from
 the honeycomb lattice in (a). This deformed lattice can be used to label the sites
 of the honeycomb lattice by column and row indices.~\label{fig1}}
\end{figure}

In the Kitaev model, a $\frac{1}{2}$-spin is placed at each site of
a honeycomb lattice [see Fig.~\ref{fig1}(a)] and the interactions
between nearest-neighbor spins are highly anisotropic with three
types of bonds $J_x,J_y$, and $J_z$. To simplify the site-labelling
of the honeycomb lattice, one can deform it to a topologically
equivalent brick-wall lattice shown in Fig.~\ref{fig1}(b). In
\cite{XYFeng,HDChen1,HDChen}, the topological quantum phase
transition of the Kitaev model on a brick-wall lattice was studied
for the Hamiltonian:
\begin{eqnarray}\label{Kitaev}
H_0&=&J_x\sum_{n+m=\text{odd}}\sigma_{n,m}^x\sigma_{n+1,m}^x\nonumber\\
&&+J_y\sum_{n+m=\text{even}}\sigma_{n,m}^y\sigma_{n+1,m}^y\nonumber\\
&&+J_z\sum_{n+m=\text{even}}\sigma_{n,m}^z\sigma_{n,m+1}^z,
\end{eqnarray}
where $\sigma_{n,m}^x,\sigma_{n,m}^y$ and $\sigma_{n,m}^z$ are the
Pauli matrices at the site $(n, m)$, with column index
$n=0,1,2,3,\cdots, N-1$ and row index $m=0,1,2,3,\cdots, M-1$. A
nice Jordan-Wigner transformation was
introduced~\cite{XYFeng,HDChen1,HDChen} to solve this model and the
redundant gauge degrees of freedom were removed.

The phase diagram of the Kitaev model (\ref{Kitaev}) consists of two
phases: A band insulator phase and a topologically non-universal
gapless phase \cite{Kitaev}. The insulator phase, as Kitaev has
shown by using perturbation theory \cite{Kitaev,vidal}, is
equivalent to a toric code model \cite{Kitaev1}. While Abelian
anyons can be defined in the insulator phase, the vortices in the
gapless phase do not have a well-defined statistics. Applying an
external magnetic field as a perturbation, which breaks the
time-reversal symmetry in Eq. (\ref{Kitaev}), a gap opens in the
gapless phase and the vortices then obey a well-defined non-Abelian
anyonic statistics \cite{Kitaev}. The third-order perturbation
corresponds to exactly soluble models~\cite{Kitaev,DHLee} whose
spectrum has recently been extensively studied \cite{Lahtinen}.

In this paper, we study the following
Hamiltonian~\cite{extendedground,DHLee,yuwang}:
\begin{eqnarray}\label{hamiltonian}
H\!&\!=\!&\!H_0+J\sum_{n+m=\text{odd}}
\sigma_{n,m}^x\sigma_{n+1,m}^z\sigma_{n+2,m}^y\nonumber\\&&\!
+J\sum_{n+m=\text{even}}\sigma_{n,m}^y\sigma_{n+1,m}^z\sigma_{n+2,m}^x.
\end{eqnarray}
Hereafter, we call the model in Eq.~(\ref{hamiltonian}) an extended
Kitaev model. We solve this model on a torus and mainly focus on the
quantum phase transition between the phase with Abelian anyons and
the phase with non-Abelian anyons. We first apply the Jordan-Wigner
transformation to the spin operators and then introduce Majorana
fermions to get the ground state of Eq.~(\ref{hamiltonian}) in the
vortex-free sector. We show that the third directional derivative of
the ground-state energy is discontinuous at each point on the
critical line separating the Abelian and non-Abelian phases, while
its first and second directional derivatives are continuous at this
point. This implies that the topological quantum phase transition is
continuous in this extended Kitaev model. Moreover, at this critical
point, we also study the nonanalyticity of the entanglement (i.e.,
the von Neumann entropy) between two nearest-neighbor spins and the
rest of the spins in the system. We find that the second directional
derivative of the von Neumann entropy is closely related to the
third directional derivative of the ground-state energy and it is
also discontinuous at the critical point. Our approach directly
reveals that both the entanglement measure and the ground-state
energy can be used to characterize the topological quantum phase
transition in the extended Kitaev model.

\begin{figure}
\includegraphics[width=3.0in,bbllx=48,bblly=542,bburx=165,bbury=653]
{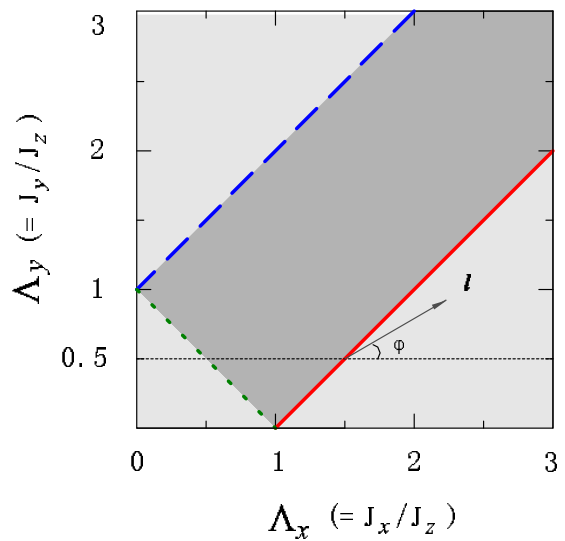}
 \caption{~\label{fig2} (Color online) Phase diagram of the extended
 Kitaev spin model, where $J_x,~J_y$ and $J_z>0$. The gray region corresponds to the non-Abelian phase and
 the three triangular (light gray) regions correspond to the Abelian
 phase. The thick solid, dashed and dotted lines are
 $\Lambda_x=1+\Lambda_y$, $\Lambda_y=1+\Lambda_x$, and
 $\Lambda_y=1-\Lambda_x$, where $\Lambda_x\equiv J_x/J_z$ and $\Lambda_y\equiv
 J_y/J_z$. These lines consist of the boundary of the gray region,
 which are the critical lines separating the Abelian and non-Abelian
 phases. The thin dotted line intersects the thick solid and dotted
 lines at the points $(\Lambda_x,~\Lambda_y)=(1.5,~0.5)$ and
 $(0.5,~0.5)$. The direction \textbf{\textit{l}} has an inclination
 angle $\varphi$ with respect to the horizontal axis and it indicates
 the direction along which the driving parameters $\Lambda_x$ and $\Lambda_y$ vary.}
\end{figure}
\section{Topological Quantum phase transition}
Let us define the Jordan-Wigner transformation~\cite{xymodel}
\begin{eqnarray}
\sigma_{n,m}^{+}&=&2\left(a_{n,m}^{(s)}\right)^{\dag}K(n,m),\nonumber\\
K(n,m)&=&\prod_{n^{\prime}=0}^{N-1}\prod_{m^{\prime}=0}^{m-1}
\sigma_{n^{\prime},m^{\prime}}^{z}
\prod_{n^{\prime}=1}^{n-1}\sigma_{n^{\prime},m}^{z},\label{phaseterm}
\end{eqnarray}
where $s=1$ if the integer $n+m$ is odd and $s=2$ if the integer
$n+m$ is even. Also, we introduce the following definitions for
Majorana fermions:
\begin{eqnarray}\label{majo001}
i\left[a_{n,m}^{(1)\dag}-a_{n,m}^{(1)}\right]&=&c_{n,m}^{(1)},\nonumber\\~a_{n,m}^{(1)\dag}+a_{n,m}^{(1)}&=&d_{n,m}^{(1)}
\end{eqnarray}
for $n+m$ equal to an odd integer, and
\begin{eqnarray}\label{majo002}
i\left[a_{n,m}^{(2)\dag}-a_{n,m}^{(2)}\right]&=&d_{n,m}^{(2)},\nonumber\\~a_{n,m}^{(2)\dag}+a_{n,m}^{(2)}&=&c_{n,m}^{(2)}
\end{eqnarray}
for $n+m$ equal to an even integer. When the phase (arising from the
Jordan-Wigner transformation) related to each bond between the
$(N-1)$th column and the zeroth column is chosen to be $2\pi l$ ($l$
is an integer ), the Hamiltonian (\ref{hamiltonian}) is reduced to
\begin{eqnarray}\label{hamilfermion}
H&=&iJ_x\sum_{n+m=\text{odd}}c_{n,m}^{(1)}c_{n+1,m}^{(2)}\nonumber\\&&
-iJ_y\sum_{n+m=\text{even}}c_{n,m}^{(2)}c_{n+1,m}^{(1)}\nonumber\\
&&+iJ_z\sum_{n+m=\text{even}}id_{n,m}^{(2)}d_{n,m+1}^{(1)}c_{n,m}^{(2)}c_{n,m+1}^{(1)}\nonumber\\
&&-iJ\sum_{n+m=\text{odd}}c_{n,m}^{(1)}c_{n+2,m}^{(1)}\nonumber\\&&+iJ\sum_{n+m=\text{even}}c_{n,m}^{(2)}c_{n+2,m}^{(2)}.
\end{eqnarray}
In Eq.~(\ref{hamilfermion}), the $\frac{1}{2}NM$ operators
$id_{n,m}^{(2)}d_{n,m+1}^{(1)}$, where $n+m$ is an even integer,
commute with each other. The ground state is in the vortex-free
sector~\cite{lieb,extendedground} with
$d_{n,m+1}^{(1)}d_{n,m}^{(2)}d_{n+2,m+1}^{(1)}d_{n+2,m}^{(2)}=-1$,
which corresponds to the case with the eigenvalue of each plaquette
operator~\cite{Kitaev}
\begin{equation}\label{plaquette}
W_{(n,m)}=\sigma_{n,m}^x\sigma_{n+1,m}^z\sigma_{n+2,m}^y\sigma_{n,m+1}^y\sigma_{n+1,m+1}^z\sigma_{n+2,m+1}^x
\end{equation}
equal to 1. Thus, we can set the $\frac{1}{2}NM$ operators
$id_{n,m}^{(2)}d_{n,m+1}^{(1)}$ all equal to 1 in
Eq.~(\ref{hamilfermion}), in order to obtain the ground-state
energy. For this quadratic Hamiltonian, the Fourier transformation
of $H$ via
$c_{n,m}^{(1)}=\frac{2}{\sqrt{NM}}\sum_{\mathbf{k}}e^{i\mathbf{k}\cdot
\mathbf{r}_{nm}}c_{\mathbf{k}}^{(1)} $ gives rise to
\begin{eqnarray}
H&=&\sum_{\mathbf{k}}\Phi_{\mathbf{k}}^{\dag}H_{\mathbf{k}}\Phi_{\mathbf{k}},\nonumber\\
H_{\mathbf{k}}&=&h_x(\mathbf{k})\sigma^x+h_y(\mathbf{k})\sigma^y+h_z(\mathbf{k})\sigma^z,\label{hamil}
\end{eqnarray}
where $\sigma^x, \sigma^y $ and $\sigma^z$ are Pauli matrices,
$\Phi_{\mathbf{k}}^{\dag}=\left(c_{\mathbf{-k}}^{(1)},c_{\mathbf{-k}}^{(2)}\right)$
with
$c_{\mathbf{-k}}^{(j)}=\left(c_{\mathbf{k}}^{(j)}\right)^{\dag}$,
and
\begin{eqnarray} \label{h2D}
h_x(\mathbf{k})&=&-J_x\sin\left(\frac{k_x}{2}+\frac{k_y}{2\sqrt{3}}\right)+J_y\sin\left(\frac{k_x}{2}-\frac{k_y}{2\sqrt{3}}\right)
\nonumber\\&&-J_z\sin\frac{k_y}{\sqrt{3}},\nonumber\\
h_y(\mathbf{k})&=&-J_x\cos\left(\frac{k_x}{2}+\frac{k_y}{2\sqrt{3}}\right)-J_y\cos\left(\frac{k_x}{2}-\frac{k_y}{2\sqrt{3}}\right)\nonumber\\
&&+J_z\cos\frac{k_y}{\sqrt{3}},\nonumber\\
h_z(\mathbf{k})&=&2J\sin k_x.
\end{eqnarray}

Let us define
\begin{eqnarray}\label{realfermions}
B_{\mathbf{k}}&=&\frac{\alpha^{\ast}(
\mathbf{k})c_{\mathbf{k}}^{(1)}-\left[\varepsilon(\mathbf{k})+2J\sin
k_x\right]c_{\mathbf{k}}^{(2)}}{\sqrt{|\alpha(
\mathbf{k})|^2+\left[\varepsilon(\mathbf{k})+2J\sin k_x\right]^2}},
\end{eqnarray}
where
\begin{equation}
\alpha( \mathbf{k})=
iJ_xe^{i\left(\frac{k_x}{2}+\frac{k_y}{2\sqrt{3}}\right)}
+iJ_ye^{-i\left(\frac{k_x}{2}-\frac{k_y}{2\sqrt{3}}\right)}-iJ_ze^{i\frac{-k_y}{\sqrt{3}}},
\end{equation}
and
\begin{eqnarray}\label{energyspectrum}
\varepsilon(\mathbf{k})&=&|\mathbf{h}(\mathbf{k})|\nonumber\\&=&\sqrt{|\alpha(
\mathbf{k})|^2+4J^2\sin^2 k_x}.
\end{eqnarray}
It is straightforward to verify that
\begin{eqnarray}
\left\{B_{\mathbf{k}}^{\dag},B_{\mathbf{k}'}\right\}=\delta_{\mathbf{k},\mathbf{k}'},
\end{eqnarray}
 i.e., $B_{\mathbf{k}}^{\dag}$ and $B_{\mathbf{k}}$ are fermionic operators, and the Hamiltonian (\ref{hamil}) can be written
as
\begin{equation}\label{realfermionhamiltonian}
H=\sum_{\mathbf{k}}\left[\varepsilon(\mathbf{k})-2\varepsilon(\mathbf{k})B_{\mathbf{k}}^{\dag}B_{\mathbf{k}}\right].
\end{equation}
For Hamiltonian (\ref{realfermionhamiltonian}), the ground-state
energy is $-\sum_{\mathbf{k}}\varepsilon(\mathbf{k})$ and the
ground-state $|g\rangle$ obeys
$B_{\mathbf{k}}^{\dag}B_{\mathbf{k}}|g\rangle=|g\rangle$. The energy
spectrum $\varepsilon(\mathbf{k})$ is {\it gapless}~\cite{DHLee}
only when $J_x=J_y+J_z$, or $J_y=J_z+J_x$, or $J_z=J_x+J_y$, which
corresponds to the thick solid, dashed, and dotted lines in Fig.
\ref{fig2}, respectively.

When $J>0$, the spectral Chern number is $1$ if $J_x<J_y+J_z,
J_y<J_z+J_z$ and $J_z<J_x+J_y$, and 0 if $J_x>J_y+J_z$ or
$J_y>J_z+J_z$ or $J_z>J_x+J_y$~(see \cite{Kitaev,DHLee}). These
two cases correspond to the non-Abelian and Abelian phases in the
 Kitaev model and both of them are {\it gapped} topological
phases. The phase diagram is shown in Fig.~\ref{fig2}, where the
gray area corresponds to the non-Abelian phase and the critical
lines (denoted as thick solid, dashed and dotted lines) separate
the Abelian and non-Abelian phases. This indicates that the system
can experience quantum phase transitions across these three thick
lines. Here we rescale the inter-spin coupling strengths by
introducing $ \Lambda_x\equiv J_x/J_z, \Lambda_y\equiv J_y/J_z$,
and $\Lambda\equiv J/J_z$, so as to conveniently characterize the
quantum phase transition.

To demonstrate the quantum phase transition, one may reveal the
nonanalyticity of the ground-state energy.
\begin{figure}
\includegraphics[width=3.1in,bbllx=182,bblly=116,bburx=437,bbury=653]
{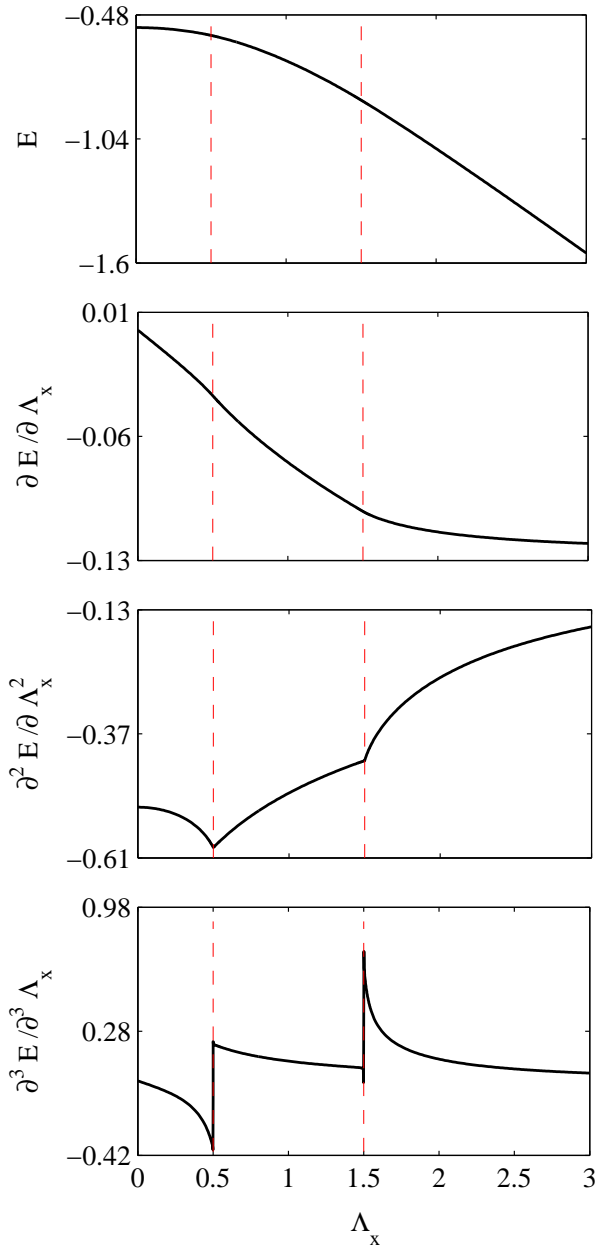} \caption{\label{fig3}The ground-state energy per site
 $E$ (in units of $J_z$) and its first, second, and
  third derivatives with respect to $\Lambda_x$, where $\Lambda_y=0.5,~\Lambda=0.1$ and $\varphi=0$
   (which corresponds to the horizontal thin dotted line in Fig.~\ref{fig2}). It is clear that
   $E$ and $\partial E/\partial \Lambda_x$ are continuous functions, but
    $\partial^3E/\partial \Lambda_x^3$ is
 discontinuous at the transition points $\Lambda_x=0.5$ and $1.5$.}
\end{figure}
The ground-state energy per site is
\begin{eqnarray}\label{groundenergy}
E=-\frac{1}{NM}\sum_{\mathbf{k}}\varepsilon(\mathbf{k})
=-\frac{\sqrt{3}}{16\pi^2}\int_{\mathrm{BZ}}d^2k\varepsilon(\mathbf{k}),
\end{eqnarray}
where $\mathrm{BZ}$ denotes the first Brillouin zone. Its
directional derivatives with respect to the driving parameter along
any given direction $\textbf{\textit{l}}$ (see Fig.~\ref{fig2}) are
\begin{eqnarray}
\frac{\partial E}{\partial l}&=&\cos \varphi\frac{\partial
E}{\partial
\Lambda_x}+\sin \varphi\frac{\partial E}{\partial \Lambda_y},\nonumber\\
\frac{\partial^2 E}{\partial l^2} &=& \cos^2 \varphi\frac{\partial^2
E}{\partial \Lambda_x^2}+\sin 2\varphi\frac{\partial^2 E}{\partial
\Lambda_x\partial \Lambda_y} +\sin^2 \varphi\frac{\partial^2E
 }{\partial \Lambda_y^2}, \nonumber\\
\frac{\partial^3 E}{\partial l^3}
 &=&\cos^3
\varphi \frac{\partial^3 E}{\partial \Lambda_x^3}+3\sin
\varphi\cos^2\varphi\frac{\partial^3 E}{\partial \Lambda_y\partial
\Lambda_x^2}  \nonumber\\&&+3\sin^2 \varphi
\cos\varphi\frac{\partial^3 E}{\partial \Lambda_x\partial
\Lambda_y^2} +\sin^3 \varphi\frac{\partial^3 E}{\partial
\Lambda_y^3},\label{directionalderivative}\nonumber\\
\cdots\!&\!\cdots\!&\!\cdots\cdots.
\end{eqnarray}
If the $n$th directional derivative $\partial^n E/\partial l^n$
($n$=1, 2, $\cdots$) is nonanalytical at the critical point
$(\Lambda_x^c, \Lambda_y^c)$, and the directional derivatives
$\partial^m E/\partial l^m$ with $0\leq m<n$ are analytical there, a
topological quantum phase transition occurs at this critical point.

 It can be proved that (see Appendix~A)
\begin{eqnarray}
 \frac{\partial  E}{\partial l }\bigg|_{1^{+}} &=&
\frac{\partial  E}{\partial l }\bigg|_{1^{-}},\nonumber\\
\frac{\partial^2 E}{\partial l^2}\bigg|_{1^{+}} &=& \frac{\partial^2
E}{\partial l^2}\bigg|_{1^{-}}  ,\label{1and2ofground}
\end{eqnarray}
and
\begin{eqnarray}
\frac{\partial^3 E}{\partial l^3}\bigg|_{1^{+}}- \frac{\partial^3
E}{\partial l^3}\bigg|_{1^{-}} & \simeq&\frac{\sqrt{6}\mathscr{D}^3
}{2\pi^2}\Gamma , \label{analyticalproveground}
\end{eqnarray}
where
\begin{eqnarray}
\mathscr{D}\!&\!=\!&\!\cos\left(\varphi+\frac{\pi}{4}\right),\nonumber\\
\Gamma\!&\!=\!&\! \int_0^{2\pi}d\theta \frac{1}{Q_1+Q_2\sin (2\theta
+\phi_1)},\label{defineformarktext}
\end{eqnarray}
$1^{+}$ denotes $(\Lambda_x-\Lambda_y)\rightarrow 1$ with
$(\Lambda_x-\Lambda_y)>1$, and $1^{-}$ denotes $
(\Lambda_x-\Lambda_y)\rightarrow1$ with $(\Lambda_x-\Lambda_y)<1$.
In (\ref{defineformarktext}),
\begin{eqnarray}
 Q_1&=&\frac{1}{2}(4\Lambda^2+\Lambda_y^2+\Lambda_y +1),
\nonumber\\
  Q_2&=&\frac{1}{4}\sqrt{\left(8\Lambda^2+2\Lambda_y^2+2\Lambda_y -1\right)^2
 + 3 (1+2\Lambda_y)^2 },\nonumber\\\label{P1andP2}
 \phi_1&=&\arctan\left(\frac{8\Lambda^2+2\Lambda_y^2+2\Lambda_y-1}{\sqrt 3
 (1+2\Lambda_y)}\right).
 \end{eqnarray}
Equations~(\ref{1and2ofground}) and (\ref{analyticalproveground})
reveal that a continuous topological quantum phase transition occurs
across the critical line $\Lambda_x=1+\Lambda_y$ (denoted by the
thick solid line in Fig.~\ref{fig2}). Similarly, it can be shown
that such a continuous topological quantum phase transition also
occurs across the critical lines $\Lambda_y=1+\Lambda_x$ and
$\Lambda_x=1-\Lambda_y$ (denoted, respectively, by the thick dashed
and dotted lines in Fig.~\ref{fig2}). As a numerical test, we choose
$\Lambda=0.1,~\varphi=0$ and $\Lambda_y=0.5$ to show this quantum
phase transition in Fig.~\ref{fig3}, where the range of $\Lambda_x$
is chosen by the thin dotted line in Fig.~\ref{fig2}. It can be seen
in Fig.~\ref{fig3} that the ground-state energy and its first and
second directional derivatives are continuous for each $\Lambda_x$,
while its third directional derivative is nonanalytic at the points
$\Lambda_x=0.5$ and $\Lambda_x=1.5$. These two points satisfy the
condition $J_z=J_x+J_y$ and $J_x=J_y+J_z$, respectively. It is
obvious that these two points are on the critical lines denoted by
the thick dotted and solid lines in Fig.~\ref{fig2}.

\section{entanglement}
It has been shown that the entanglement also exhibits critical
behavior at the quantum phase transition point for both spin (see,
e.g., \cite{entanforphase01,entanforphase02,entanforphase03}) and
fermionic systems~(see, e.g.,
\cite{entanforphase04,entanforphase05}). Also, it was
shown~\cite{laWu02,laWu01} that there is a general relation between
the bipartite entanglement and the quantum phase transition. In this
section, we show that the nonanalyticity of the ground-state energy
in the extended Kitaev model results from the correlation functions
[see Eqs.~(\ref{x-bond density}), (\ref{y-bond density}) and
(\ref{z-bond density}) for their definitions]. Furthermore, we show
that the bipartite entanglement also exhibits nonanalyticity at the
quantum phase transition point and its nonanalyticity is also due to
the nonanalyticity of the same correlation functions. This reveals
that both the ground-state energy and the bipartite entanglement can
characterize the quantum phase transition in the Kitaev model.
\subsection{Correlation functions and nonanalyticity of ground-state energy }

 From
Hellmann-Feynman theorem~\cite{hellmann}, we have
\begin{eqnarray}\label{HF}
\frac{\partial E}{\partial
l}\!&\!=\!&\!\frac{1}{NM}\mathrm{Tr}\left( \rho\frac{\partial
H}{\partial l}\right)\nonumber\\\!&\!=\!&\!\frac{1}{NM}\cos\varphi\sum_{n+m=\text{odd}}\mathrm{Tr}\left( \rho\sigma_{n,m}^x\sigma_{n+1,m}^x\right)\nonumber\\
&&\!+\frac{1}{NM}\sin\varphi\sum_{n+m=\text{even}}\mathrm{Tr}\left(
\rho\sigma_{n,m}^y\sigma_{n+1,m}^y\right),
\end{eqnarray}
where $E$ is the ground-state energy per site given in
Eq.~(\ref{groundenergy}), $H$ is the Hamiltonian (\ref{hamiltonian})
(rescaled by $J_z$), $\mathrm{Tr}$ denotes the trace over the
ground-state subspace, and $ \rho=|g\rangle\langle g|$ is the
density matrix of the system.  When $|g\rangle\langle g|$ is traced
over all spins except the two spins at $\mathbf{r}_{n,m}$ and
$\mathbf{r}_{n',m'}$, the reduced density matrix is
\begin{eqnarray}\label{densitymatrix}
\rho(\mathbf{r}_{n,m},\mathbf{r}_{n',m'})&=&\mathrm{Tr}'\left(|g\rangle\langle g|\right)\nonumber\\
&=&\frac{1}{4}\sum_{\alpha,\alpha'=0}^3 \left\langle
g\left|\sigma_{{n,m}}^{\alpha}\sigma_{{n',m'}}^{\alpha'}\right|g\right\rangle\nonumber\\&&\times
\sigma_{{n,m}}^{\alpha}\sigma_{{n',m'}}^{\alpha'},
\end{eqnarray}
where $\sigma^{\alpha}$$($$\sigma^{\alpha'}$$)$ are Pauli matrices
$\sigma^{x}, \sigma^{y}$ and $ \sigma^{z}$ for $\alpha$
$($$\alpha'$$)$$=1$ to 3, and the unit matrix for $\alpha$
$($$\alpha'$$)$$=0$. When the two spins at $\mathbf{r}_{n,m}$ and
$\mathbf{r}_{n+1,m}$ are linked by an $x$-type bond, the reduced
density matrix becomes
\begin{eqnarray}
\rho(\mathbf{r}_{n,m},\mathbf{r}_{n+1,m})\!&\!=\!&\!\frac{1}{4}\left\langle
g\left|\sigma_{{n,m}}^x\sigma_{{n+1,m}}^x\right|g\right\rangle
\sigma_{{n,m}}^x\sigma_{{n+1,m}}^x\nonumber\\&&\!+\frac{1}{4}I_{{n,m}}
I_{{n+1,m}}\label{xbond00}
\end{eqnarray}
where $n+m$ is an odd integer, and $I$ is the unit operator.

 Because of translational
invariance, the correlation function
\begin{equation}
\mathcal {G}_x\equiv\left\langle
g\left|\sigma_{{n,m}}^x\sigma_{{n+1,m}}^x\right|g\right\rangle\label{x-bond
density}
\end{equation}
 is spatially invariant. Thus,
Eq.~(\ref{xbond00}) can be written as
\begin{eqnarray}
\rho(\mathbf{r}_{n,m},\mathbf{r}_{n+1,m})\!&\!=\!&\!\frac{I_{{n,m}}
I_{{n+1,m}}+\mathcal
{G}_x\sigma_{{n,m}}^x\sigma_{{n+1,m}}^x}{4},\label{x-bond}
\end{eqnarray}
where $n+m$ is an odd integer. Similarly, one has
\begin{eqnarray}
\rho(\mathbf{r}_{n,m},\mathbf{r}_{n+1,m})\!&\!=\!&\!\frac{I_{{n,m}}
I_{{n+1,m}}+\mathcal {G}_y\sigma_{{n,m}}^y\sigma_{{n+1,m}}^y}{4},\label{y-bond}\\
\rho(\mathbf{r}_{n,m},\mathbf{r}_{n,m+1})\!&\!=\!&\!\frac{I_{{n,m}}
I_{{n,m+1}}+\mathcal
{G}_z\sigma_{{n,m}}^z\sigma_{{n,m+1}}^z}{4},\label{z-bond}
\end{eqnarray}
with
\begin{eqnarray}
\mathcal{G}_y&\equiv&\left\langle
g\left|\sigma_{{n,m}}^y\sigma_{{n+1,m}}^y\right|g\right\rangle,\label{y-bond density}\\
\mathcal {G}_z&\equiv&\left\langle
g\left|\sigma_{{n,m}}^z\sigma_{{n,m+1}}^z\right|g\right\rangle,\label{z-bond
density}
\end{eqnarray}
where $n+m$ is an even integer for both $\mathcal {G}_y$ and
$\mathcal {G}_z$. Here Eqs.~(\ref{x-bond})-(\ref{z-bond}) are the
results obtained for the reduced density matrix when the two spins
at $\mathbf{r}_{n,m}$ and $\mathbf{r}_{n',m'}$ are
nearest-neighbors. When the two spins at $\mathbf{r}_{n,m}$ and
$\mathbf{r}_{n',m'}$ are not nearest-neighbors, the density matrix
is
\begin{eqnarray}\rho(\mathbf{r}_{n,m},\mathbf{r}_{n',m'})=\frac{I_{{n,m}}
I_{{n',m'}}}{4}.
\end{eqnarray}

Using the Jordan-Wigner
transformation~(\ref{phaseterm}) and the definitions~(\ref{majo001})
and (\ref{majo002}) for the Majorana fermions, we
 can derive that
\begin{eqnarray}
\mathcal {G}_{ x}=\left\langle
g\left|\sigma_{0,1}^{x}\sigma_{1,1}^{x}\right|g\right\rangle=
i\left\langle g\left|
c_{0,1}^{(1)}c_{1,1}^{(2)}\right|g\right\rangle.\label{howtocorre}
\end{eqnarray}
From Eqs.~(\ref{realfermions}), (\ref{energyspectrum}), and
(\ref{howtocorre}) we have
\begin{eqnarray}\label{correlation001}
\mathcal
{G}_x\!&\!=\!&\!\frac{\sqrt{3}i}{8\pi^2}\int_{\mathrm{BZ}}d^2k\frac{\alpha(
\mathbf{k})}{\varepsilon(\mathbf{k})}\mathrm{exp}\left(-i\frac{k_x}{2}-
i\frac{\sqrt{3}k_y}{6} \right)\nonumber\\
\!&\!=\!&\!\frac{-\sqrt{3}}{8\pi^2}\int_{\mathrm{BZ}}d^2k
\frac{1}{\sqrt{|\alpha(
\mathbf{k})|^2+4\Lambda^2\sin^2k_x}}\nonumber\\&&\times
\left[\Lambda_x+\Lambda_y\cos k_x-\cos
\frac{k_x+\sqrt{3}k_y}{2}\right],
\end{eqnarray}
and
\begin{eqnarray}
|\mathcal
{G}_x|&\leq&\frac{\sqrt{3}}{8\pi^2}\int_{\mathrm{BZ}}d^2k\frac{|\alpha(
\mathbf{k})|}{\varepsilon(\mathbf{k})}\nonumber\\
&<&\frac{\sqrt{3}}{8\pi^2}\int_{\mathrm{BZ}}d^2k\frac{\varepsilon(\mathbf{k})}{\varepsilon(\mathbf{k})}=1,\label{rangeofG}
\end{eqnarray}
which gives rise to $ -1<\mathcal {G}_x<1$. Similarly, we have
\begin{eqnarray}\label{correlation002}
\mathcal {G}_y
\!&\!=\!&\!\frac{-\sqrt{3}}{8\pi^2}\int_{\mathrm{BZ}}d^2k
\frac{1}{\sqrt{|\alpha(
\mathbf{k})|^2+4\Lambda^2\sin^2k_x}}\nonumber\\&&\times
\left[\Lambda_x\cos k_x+\Lambda_y-\cos
\frac{k_x-\sqrt{3}k_y}{2}\right],
\end{eqnarray}
with $ -1<\mathcal {G}_y<1$.

From Eqs.~(\ref{HF}), (\ref{x-bond density}), and (\ref{y-bond
density}), it follows that
\begin{eqnarray}
\frac{\partial E}{\partial l}
&=&\frac{1}{2}\cos\varphi\mathcal{G}_x+\frac{1}{2}\sin\varphi\mathcal{G}_y,\nonumber\\
\frac{\partial^2 E}{\partial l^2}
&=&\frac{1}{2}\cos^2\varphi\frac{\partial
\mathcal{G}_x}{\partial\Lambda_x}+\frac{1}{2}\sin^2\varphi\frac{\partial
\mathcal{G}_y}{\partial\Lambda_y}\nonumber\\
&&+\frac{1}{2}\sin\varphi\cos\varphi\left[\frac{\partial
\mathcal{G}_x}{\partial\Lambda_y}+\frac{\partial
\mathcal{G}_y}{\partial\Lambda_x}\right],
\nonumber\\
 \frac{\partial^3 E}{\partial l^3}
&=&\frac{1}{2}\cos^3\varphi\frac{\partial^2
\mathcal{G}_x}{\partial\Lambda_x^2}
+\frac{1}{2}\sin^3\varphi\frac{\partial^2
\mathcal{G}_y}{\partial\Lambda_y^2}\nonumber\\&&+\frac{1}{2}\cos^2\varphi\sin\varphi\left[2\frac{\partial^2
\mathcal{G}_x}{\partial\Lambda_x\partial\Lambda_y}+\frac{\partial^2
\mathcal{G}_y}{\partial\Lambda_x^2}\right]\nonumber\\&&+\frac{1}{2}\cos\varphi\sin^2\varphi\left[2\frac{\partial^2
\mathcal{G}_y}{\partial\Lambda_x\partial\Lambda_y}+\frac{\partial^2
\mathcal{G}_x}{\partial\Lambda_y^2}\right].\label{correforground}
\end{eqnarray}
Equation (\ref{correforground}) shows that the directional
derivatives of the ground-state energy per site are determined by
the correlation functions $\mathcal{G}_{\alpha}$ ($\alpha=x,~y$) and
their derivatives. Section II shows that $\partial E/\partial l$ and
$\partial^2 E/\partial l^2$ are continuous, while $\partial^3
E/\partial l^3$ is \textit{discontinuous} on the critical line,
e.g., $\Lambda_x=1+\Lambda_y$ (i.e., the thick solid line in
Fig.~\ref{fig2}). Equation~(\ref{correforground}) reveals that the
nonanalyticity of $E$ on the critical line is due to the
nonanalyticity of $\mathcal{G}_{\alpha}$. As shown in Appendix~B,
\begin{eqnarray}
\mathcal{G}_{\alpha}\big|_{1^+}\!&\!=\!&\!\mathcal{G}_{\beta}\big|_{1^-},\nonumber\\
\frac{\partial\mathcal{G}_{\alpha}}{\partial
\Lambda_{\beta}}\bigg|_{1^+}\!&\!=\!&\!\frac{\partial\mathcal{G}_{\alpha}}{\partial
\Lambda_{\beta}}\bigg|_{1^-},\label{contcorrtext}
\end{eqnarray}
where $\alpha,~\beta=x,~y$, and
\begin{eqnarray}
 \frac{\partial^2 \mathcal{G}_{x}}{\partial \Lambda_x^2}\bigg|_{1^+}-\frac{\partial^2 \mathcal{G}_{x}}{\partial
 \Lambda_x^2}\bigg|_{1^-}\!&\!\simeq\!&\!
\frac{\sqrt{3} }{2\pi^2}\Gamma,\nonumber\\
 \frac{\partial^2 \mathcal{G}_{x}}{\partial \Lambda_x\partial \Lambda_y}\bigg|_{1^+}
-\frac{\partial^2 \mathcal{G}_{x}}{\partial \Lambda_x\partial
\Lambda_y}\bigg|_{1^-}\!&\!\simeq\!&\!
-\frac{\sqrt{3} }{2\pi^2}\Gamma,\nonumber\\
 \frac{\partial^2 \mathcal{G}_{x}}{\partial \Lambda_y^2}\bigg|_{1^+}-\frac{\partial^2 \mathcal{G}_{x}}{\partial
 \Lambda_y^2}\bigg|_{1^-}\!&\!\simeq\!&\!
\frac{\sqrt{3} }{2\pi^2}\Gamma,\nonumber\\
 \frac{\partial^2 \mathcal{G}_{y}}{\partial \Lambda_x^2}\bigg|_{1^+}-\frac{\partial^2 \mathcal{G}_{y}}{\partial
 \Lambda_x^2}\bigg|_{1^-}\!&\!\simeq\!&\!
-\frac{\sqrt{3} }{2\pi^2}\Gamma,\nonumber\\
 \frac{\partial^2 \mathcal{G}_{y}}{\partial \Lambda_x\partial \Lambda_y}\bigg|_{1^+}
-\frac{\partial^2 \mathcal{G}_{y}}{\partial \Lambda_x\partial
\Lambda_y}\bigg|_{1^-}\!&\!\simeq\!&\!
\frac{\sqrt{3} }{2\pi^2}\Gamma,\nonumber\\
 \frac{\partial^2 \mathcal{G}_{y}}{\partial \Lambda_y^2}\bigg|_{1^+}-\frac{\partial^2 \mathcal{G}_{y}}{\partial
 \Lambda_y^2}\bigg|_{1^-}\!&\!\simeq\!&\!
-\frac{\sqrt{3} }{2\pi^2}\Gamma,\label{disofthecorrtext}
\end{eqnarray}
where $\Gamma$ is given in Eq.~(\ref{defineformarktext}). Equation
(\ref{disofthecorrtext}) shows that the spin-spin correlation
function $\mathcal{G}_{\alpha}$ can signal the quantum phase
transition, similar to the bond-bond correlation function in the
original Kitaev model~\cite{Suncp}. From Eqs.~(\ref{correforground})
and (\ref{disofthecorrtext}), we have
\begin{eqnarray}
\frac{\partial^3 E}{\partial l^3}\bigg|_{1^{+}}- \frac{\partial^3
E}{\partial l^3}\bigg|_{1^{-}} &
\simeq&\frac{\sqrt{3}(\cos\varphi-\sin\varphi)^3 }{4\pi^2}\Gamma
=\frac{\sqrt{6}\mathscr{D}^3 }{2\pi^2}\Gamma,\nonumber\\
\end{eqnarray}
which is the same as in Eq.~(\ref{analyticalproveground}). This
further reveals that the nonanalyticity of the ground-state energy
results from the nonanalyticity of the correlation functions
$\mathcal{G}_{\alpha}$.

\subsection{Nonanalyticity of entanglement}
We now focus on the bipartite entanglement of the ground state
$|g\rangle$ between two spins (at $\mathbf{r}_{n,m}$ and
$\mathbf{r}_{n',m'}$) and the rest of the spins in the system. We
use the von Neumann entropy to measure the entanglement between
these two spins and the rest of the spins in the system. The von
Neumann entropy can be defined by~\cite{Nielsen}
\begin{eqnarray}
S_{\alpha}=-\text{Tr}\left[\rho(\mathbf{r}_{n,m},\mathbf{r}_{n+1,m})\log_2\rho(\mathbf{r}_{n,m},\mathbf{r}_{n+1,m})\right],
\end{eqnarray}
where \textrm{Tr} denotes the trace over the two-spin Hilbert space,
and $\alpha=x$ if $n+m$ is an odd integer, and $\alpha=y$ if $n+m$
is an even integer. Also, this entropy can be written as
\begin{equation}S_{\alpha}=-\sum\lambda_i\log_2\lambda_i,
\end{equation}
where the sum runs over the four eigenvalues $\lambda_i$ of the
matrix $\rho(\mathbf{r}_{n,m},\mathbf{r}_{n+1,m})$. From
Eqs.~(\ref{x-bond density}) and (\ref{y-bond density}), it follows
that
\begin{eqnarray}
\lambda_1\!&\!=\!&\!\lambda_2=\frac{1}{4}(1+\mathcal {G}_{\alpha}),\nonumber\\
\lambda_3\!&\!=\!&\!\lambda_4=\frac{1}{4}(1-\mathcal {G}_{\alpha}).
\end{eqnarray}
Thus, we have the entanglement measure
\begin{eqnarray}\label{entropy}
S_{\alpha}&=&2-\frac{1}{2}
\log_2\left[(1-\mathcal{G}_{\alpha})^{1-\mathcal{G}_{\alpha}}(1+\mathcal{G}_{\alpha})^{1+\mathcal{G}_{\alpha}}\right],
\end{eqnarray}
which is determined by the correlation function
$\mathcal{G}_{\alpha}$, similar to the thermal
entanglement~\cite{xgwang}.

\begin{figure}
\includegraphics[width=3.1in,bbllx=167,bblly=150,bburx=473,bbury=622]
{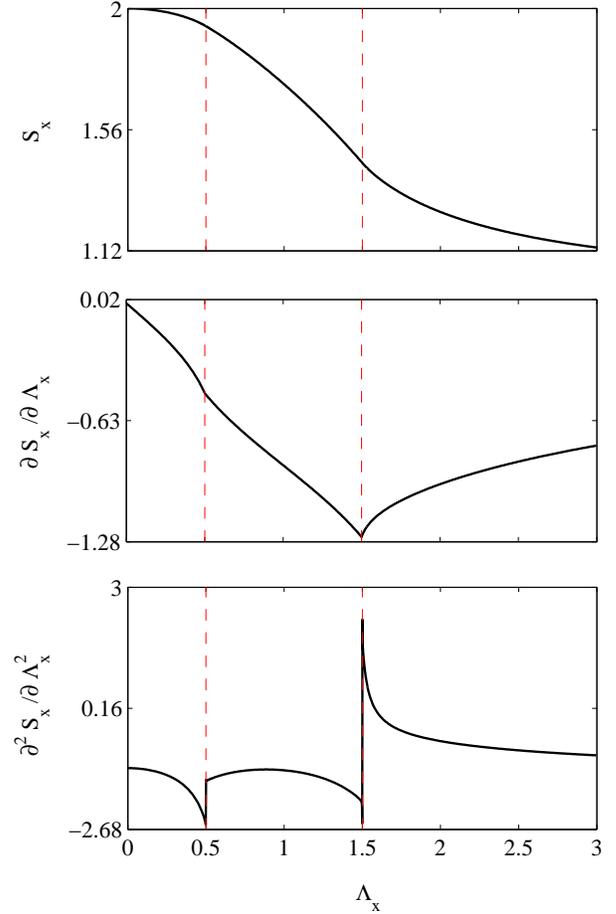}
 \caption{\label{fig4} The bipartite entanglement (i.e., the von Neumann entropy)
  $S_x$ and its first, and second
  derivatives with respect to $\Lambda_x$, where $\Lambda_y=0.5,~\Lambda=0.1$ and $\varphi=0$
   (which corresponds to the horizontal thin dotted line in Fig.~\ref{fig2}). Obviously,
   $S_x$ and $\partial S_x/\partial \Lambda_x$ are
  continuous functions, but $\partial^2 S_x/\partial \Lambda_x^2$ is
 discontinuous at the transition points $\Lambda_x=0.5$ and $1.5$.}
\end{figure}

To see the relationship between the entanglement and the quantum
phase transition, we analyze the directional derivatives of the von
Neumann entropy with respect to the driving parameters along any
direction $\textbf{\textit{l}}$. The first and second directional
derivatives of the bipartite entanglement are
\begin{eqnarray}
\frac{\partial S_{\alpha}}{\partial l}&=&\frac{\partial \mathcal
{G}_{\alpha}}{\partial l}\log_2\sqrt{\frac{1-\mathcal
{G}_{\alpha}}{1+\mathcal {G}_{\alpha}}},
\nonumber\\
\frac{\partial^2 S_{\alpha}}{\partial l^2}&=&\frac{\partial^2
\mathcal {G}_{\alpha}}{\partial l^2}\log_2\sqrt{\frac{1-\mathcal
{G}_{\alpha} }{1+\mathcal {G}_{\alpha}}} -\frac{1}{\ln
2}\frac{\partial \mathcal {G}_{\alpha}}{\partial
l}\frac{1}{1-\mathcal {G}_{\alpha}^2},\label{derivative1and2}
\end{eqnarray}
where
\begin{eqnarray}
\frac{\partial \mathcal {G}_{\alpha}}{\partial
l}&=&\cos\varphi\frac{\partial \mathcal {G}_{\alpha}}{\partial
\Lambda_x}+ \sin\varphi\frac{\partial \mathcal
{G}_{\alpha}}{\partial \Lambda_y},
\nonumber\\
\frac{\partial^2 \mathcal{G}_{\alpha}}{\partial
l^2}&=&\cos^2\varphi\frac{\partial^2
\mathcal{G}_{\alpha}}{\partial\Lambda_x^2}+\sin2\varphi\frac{\partial^2
\mathcal{G}_{\alpha}}{\partial\Lambda_x\partial\Lambda_y}+\sin^2\varphi\frac{\partial^2
\mathcal{G}_{\alpha}}{\partial\Lambda_y^2}.\nonumber\\
\end{eqnarray}

 From
Eqs.~(\ref{contcorrtext}), (\ref{disofthecorrtext}),
(\ref{entropy}),  and (\ref{derivative1and2}), we have
\begin{eqnarray}
 S_{\alpha}\big|_{1^{+}}\!&\!=\!&\!  S_{\alpha}\big|_{1^{-}},\nonumber\\
\frac{\partial S_{\alpha}}{\partial l }\bigg|_{1^{+}}
\!&\!=\!&\!\frac{\partial S_{\alpha}}{\partial l }\bigg|_{1^{-}} ,
\end{eqnarray}
and
\begin{eqnarray}
 \frac{\partial^2 S_x}{\partial l^2}\bigg|_{1^{+}}-
\frac{\partial^2 S_x}{\partial l^2}\bigg|_{1^{-}}
 \!&\!\simeq\!&\!\frac{\sqrt{3}\Gamma\mathscr{D}^2 }{\pi^2}\log_2\sqrt{\frac{1-\mathcal
{G}_{x}^c }{1+\mathcal {G}_{x}^c}} , \nonumber\\ \label{anavon2}
\frac{\partial^2 S_y}{\partial l^2}\bigg|_{1^{+}}- \frac{\partial^2
S_y}{\partial l^2}\bigg|_{1^{-}}
 \!&\!\simeq\!&\!-\frac{\sqrt{3}\Gamma\mathscr{D}^2 }{\pi^2}\log_2\sqrt{\frac{1-\mathcal
{G}_{y}^c }{1+\mathcal {G}_{y}^c}} ,
\end{eqnarray}
where
\begin{eqnarray}
\mathcal {G}_{\alpha}^c=\mathcal {G}_{\alpha}\big|_{1^+}=\mathcal
{G}_{\alpha}\big|_{1^-}.
\end{eqnarray}
Equation (\ref{anavon2}) shows that the bipartite entanglement is
nonanalytic with its second directional derivative $\partial^2
S_{\alpha}/\partial l^2$ discontinuous at the critical line
$\Lambda_x=1+\Lambda_y$ (denoted by the thick solid line in
Fig.~\ref{fig2}). Because $-1<\mathcal {G}_{\alpha}<1$, it follows
from Eq.~(\ref{derivative1and2}) that the discontinuity of
$\partial^2 S_{\alpha}/\partial l^2$ is due to the discontinuity of
$\partial^2 \mathcal {G}_{\alpha}/\partial l^2$. Similarly, it can
be shown that $\partial^2 S_{\alpha}/\partial l^2$ also exhibits a
discontinuity on the critical lines $\Lambda_y=1+\Lambda_x$ and
$\Lambda_x=1-\Lambda_y$ (denoted by the thick dashed and dotted
lines in Fig.~\ref{fig2}) which is due to the discontinuity of
$\partial^2 \mathcal{G}_{\alpha}/\partial l^2$ on these lines. As in
Fig.~\ref{fig3}, we choose $\Lambda=0.1,~\varphi=0$ and
$\Lambda_y=0.5$ as a typical example to show $S_{x}$ and its first
and second derivatives with respect to $\Lambda_x$ (see
Fig.~\ref{fig4}). It is clear that $S_{x}$ and its first derivative
$\partial S_{x}/\partial \Lambda_x$ are continuous as a function of
$\Lambda_x$, but its second derivative $\partial^2 S_{x}/\partial
\Lambda_x^2$ is discontinuous at the quantum phase transition points
$\Lambda_x=0.5$ and $\Lambda_x=1.5$.

As shown above, both the nonanalyticity of the ground-state energy
and that of the bipartite entanglement are due to the nonanalyticity
of the spin-spin correlation functions. This reveals that the
ground-state energy and the bipartite entanglement are closely
related with each other, and both of them can be used to
characterize the topological quantum phase transition in the
extended Kitaev spin model.

\section{conclusion}
In conclusion, we have studied the topological quantum phase
transition between Abelian and non-Abelian phases in the extended
Kitaev spin model on a honeycomb lattice. From the ground-state
energy, we show that this model displays a continuous quantum phase
transition on the critical lines separating the Abelian and
non-Abelian phases, where the third derivative of the ground-state
energy is discontinuous. Also, we use the von Neumann entropy as a
measure of bipartite entanglement to study this topological quantum
phase transition. Our results show that the bipartite entanglement
is also nonanalytic on the same critical lines as the ground-state
energy. Moreover, we show that the discontinuity of the second
derivative of the bipartite entanglement is related to the
discontinuity of the third derivative of the ground-state energy.
Our approach directly reveals that both the entanglement and the
ground-state energy can be used to characterize the topological
quantum phase transition in this Kitaev model.

\begin{acknowledgments}
We thank Z. D. Wang and Y. Chen for useful discussions. J.Q.Y. and
X.F.S. were supported in part by the National Basic Research Program
of China grant Nos. 2009CB929300 and 2006CB921205, the National
Natural Science Foundation of China grant Nos. 10625416, and the
MOST International Collaboration Program grant No. 2008DFA01930.
Y.Y. was supported in part by the National Natural Science
Foundation of China, the National Basic Research Program of China
and a fund from CAS. F.N. was supported in part by the U.S. National
Security Agency, the Laboratory for Physical Sciences, the U.S. Army
Research Office, and the National Science Foundation Grant No.
EIA-0130383.
\end{acknowledgments}

\appendix
\section{}
This appendix focuses on the analyticity of the first, second and
third directional derivatives of the ground-state energy on the
critical line denoted by the thick solid line in Fig.~\ref{fig2}.
From Eq.~(\ref{groundenergy}), the derivative of the ground-state
energy with respect to $\Lambda_x$ is
\begin{eqnarray}
 \frac{\partial E}{\partial \Lambda_x}&=&-\frac{\sqrt{3}}{16\pi^2}\int_{\mathrm{BZ}}d^2k\frac{\partial }{\partial
\Lambda_x}\varepsilon(\mathbf{k})\nonumber\\
&=&-\frac{\sqrt{3}}{16\pi^2}\int_{\mathrm{BZ}}d^2k\frac{A}{\varepsilon(\mathbf{k})}\nonumber\\
&=&-\frac{\sqrt{3}}{16\pi^2}\int_{\mathrm{D}}d^2k\frac{A}{\varepsilon(\mathbf{k})}
-\frac{\sqrt{3}}{16\pi^2}\int_{\mathrm{BZ-D}}d^2k\frac{A}{\varepsilon(\mathbf{k})},\label{A1}\nonumber\\
\end{eqnarray}
where
\begin{eqnarray}A=\Lambda_x+\Lambda_y\cos
k_x-\cos\left(\frac{k_x+\sqrt 3 k_y}{2}\right),\label{A2}
\end{eqnarray}
and $\mathrm{D}$ denotes two small regions in the first Brillioun
zone, i.e., half of the disk with radius $\epsilon$, which is
centered at $(\pi,~-\pi/\sqrt 3)$, and half of the disk with radius
$\epsilon$, which is centered at $(-\pi,~\pi/\sqrt 3)$, where
$\epsilon \ll 1$. When $\Lambda_x-\Lambda_y=1$,
$\varepsilon(\mathbf{k})$ becomes zero
 only at the points
$(\mp\pi,~\pm\pi/\sqrt 3)$, so
$\int_{\mathrm{BZ-D}}d^2k\frac{A}{\varepsilon(\mathbf{k})}$ is
analytic because $\mathrm{BZ-D}$ is the region excluding
$\mathrm{D}$ in the first Brillioun zone. For the integral in the
region $\mathrm{D}$, we can approximate it as
\begin{eqnarray}
 \int_{\mathrm{D}}d^2k\frac{A}{\varepsilon(\mathbf{k})}
&\simeq&\int_0^{2\pi
}d\theta\int_0^{\epsilon}KdK\frac{\Delta+K^2S_1}{\sqrt{\Delta^2
+K^2S_2}}\nonumber\\
&=&
  \int_0^{2\pi}d\theta
\left[\sqrt{ \Delta^2+\epsilon^2S_2}
-|\Delta|\right]\frac{\Delta}{S_2}\nonumber\\
&&+\int_0^{2\pi}d\theta
\left[\frac{1}{3}\left(\Delta^2+\epsilon^2S_2\right)^{\frac{3}{2}}-|\Delta|^3\right]\frac{S_1} {S_2^2}\nonumber\\
&&-\int_0^{2\pi}d\theta
\frac{S_1\Delta^2\sqrt{\Delta^2+\epsilon^2S_2}} {S_2^2}
\nonumber\\
&&+\int_0^{2\pi}d\theta|\Delta|^3\frac{S_1} {S_2^2},\label{A3}
\end{eqnarray}
where
\begin{eqnarray}\label{defineformark}
\Delta&=&\Lambda_x-\Lambda_y-1,\nonumber\\
 S_1&=&P_1+P_2\sin
(2\theta+\phi_2),\nonumber\\
 S_2&=&Q_1+Q_2\sin(2\theta+\phi_1),
 \end{eqnarray}
 and
 \begin{eqnarray}
 Q_1&=&\frac{\Lambda_x\Lambda_y+4\Lambda^2+\Lambda_x-\Lambda_y}{2},\nonumber\\
 Q_2&=&\frac{1}{4}\sqrt{\left(2\Lambda_x\Lambda_y+8\Lambda^2-\Lambda_x+\Lambda_y\right)^2
 + 3 (\Lambda_x+\Lambda_y)^2},
\nonumber\\P_1&=&\frac{\Lambda_y+1}{4},\nonumber\\P_2&=&\frac{1}{4}\sqrt{\Lambda_y^2-\Lambda_y+1},\nonumber\\
 \phi_1&=&\arctan\left(\frac{2\Lambda_x\Lambda_y+8\Lambda^2-\Lambda_x+\Lambda_y}{\sqrt 3 (\Lambda_x+\Lambda_y)}\right),\nonumber\\
 \phi_2&=& \arctan\left(\frac{2\Lambda_y-1}{\sqrt 3}\right).
 \end{eqnarray}
From Eq.~(\ref{A3}), we have
\begin{eqnarray}
\int_{\mathrm{D}}d^2k\frac{A}{\varepsilon(\mathbf{k})}\bigg|_{1^+}
&\simeq&\frac{\epsilon^3}{3}\int_0^{2\pi}d\theta\frac{S_1}
{S_2^2},\nonumber\\
\int_{\mathrm{D}}d^2k\frac{A}{\varepsilon(\mathbf{k})}\bigg|_{1^-}
&\simeq&\frac{\epsilon^3}{3}\int_0^{2\pi}d\theta\frac{S_1}
{S_2^2},\label{A5}
\end{eqnarray}
where $1^+$ denotes $\Lambda_x-\Lambda_y\rightarrow1$ with
$\Lambda_x-\Lambda_y>1$, and $1^-$ denotes
$\Lambda_x-\Lambda_y\rightarrow1$ with $\Lambda_x-\Lambda_y<1$. When
$\epsilon\rightarrow 0$, it follows from Eq.~(\ref{A5}) that
$\int_{\mathrm{BZ-D}}d^2k\frac{A}{\varepsilon(\mathbf{k})}\rightarrow
0$ on the critical line $\Lambda_x=1+\Lambda_y$ (i.e., the thick
solid line in Fig.~\ref{fig2}). Thus, from Eq.~(\ref{A1}), we have
\begin{eqnarray}
 \frac{\partial E}{\partial \Lambda_x}\bigg|_{1^+}=\frac{\partial E}{\partial
 \Lambda_x}\bigg|_{1^-}.\label{A6}
\end{eqnarray}
Similarly,
\begin{eqnarray}
 \frac{\partial E}{\partial \Lambda_y}\bigg|_{1^+}=\frac{\partial E}{\partial
 \Lambda_y}\bigg|_{1^-}.\label{A7}
\end{eqnarray}
From Eqs.~(\ref{directionalderivative}), (\ref{A6}), and (\ref{A7}),
it follows that
\begin{eqnarray}
 \frac{\partial E}{\partial l}\bigg|_{1^+}=\frac{\partial E}{\partial
 l}\bigg|_{1^-}.\label{A8}
\end{eqnarray}
Using the same procedure as above, we can obtain
\begin{eqnarray}
\frac{\partial^2 E}{\partial
l^2}\bigg|_{1^+}\!&\!=\!&\!\frac{\partial^2 E}{\partial
 l^2}\bigg|_{1^-},\nonumber\\
 \frac{\partial^3 E}{\partial l^3}\bigg|_{1^+}-\frac{\partial^3
E}{\partial
 l^3}\bigg|_{1^-}\!&\!\simeq\!&\!
\frac{\sqrt{3}\Gamma}{4\pi^2}(\cos\varphi-\sin\varphi)^3,\label{A9}
\end{eqnarray}
where
\begin{eqnarray}
\Gamma\equiv \int_0^{2\pi}d\theta \frac{1}{Q_1+Q_2\sin
(2\theta+\phi_1)}.\label{A10}
\end{eqnarray}

\section{}
This appendix gives results regarding the analyticity of the
correlation function $\mathcal{G}_{\alpha}$ ($\alpha=x,~y$) and its
first, and second directional derivatives on the critical line
denoted by the thick solid line in Fig.~\ref{fig2}. Similar to
Eq.~(\ref{A1}), one can divide the integral in
(\ref{correlation001}) into two parts:
\begin{eqnarray}
\mathcal{G}_{x}
\!&\!=\!&\!-\frac{\sqrt{3}}{8\pi^2}\int_{\mathrm{BZ}}d^2k\frac{A}{\varepsilon(\mathbf{k})}\nonumber\\
\!&\!=\!&\!-\frac{\sqrt{3}}{8\pi^2}\int_{\mathrm{D}}d^2k\frac{A}{\varepsilon(\mathbf{k})}
-\frac{\sqrt{3}}{8\pi^2}\int_{\mathrm{BZ-D}}d^2k\frac{A}{\varepsilon(\mathbf{k})},\label{B1}
\end{eqnarray}
where $A$ is given in Eq.~(\ref{A2}). Using the same procedure for
Eqs.~(\ref{A8}) and (\ref{A9}), we can derive from Eq.~(\ref{B1})
that
\begin{eqnarray}
\mathcal{G}_{x}\big|_{1^+}=\mathcal{G}_{x}\big|_{1^-},
\end{eqnarray}
and
\begin{eqnarray}
 \frac{\partial \mathcal{G}_{x}}{\partial \Lambda_x}\bigg|_{1^+}\!&\!=\!&\!\frac{\partial \mathcal{G}_{x}}{\partial
 \Lambda_x}\bigg|_{1^-},\nonumber\\
 \frac{\partial \mathcal{G}_{x}}{\partial \Lambda_y}\bigg|_{1^+}\!&\!=\!&\!\frac{\partial \mathcal{G}_{x}}{\partial
 \Lambda_y}\bigg|_{1^-},\nonumber\\
 \frac{\partial^2 \mathcal{G}_{x}}{\partial \Lambda_x^2}\bigg|_{1^+}-\frac{\partial^2 \mathcal{G}_{x}}{\partial
 \Lambda_x^2}\bigg|_{1^-}\!&\!\simeq\!&\!
\frac{\sqrt{3} }{2\pi^2}\Gamma,\nonumber\\
 \frac{\partial^2 \mathcal{G}_{x}}{\partial \Lambda_x\partial \Lambda_y}\bigg|_{1^+}
-\frac{\partial^2 \mathcal{G}_{x}}{\partial \Lambda_x\partial
\Lambda_y}\bigg|_{1^-}\!&\!\simeq\!&\!
-\frac{\sqrt{3} }{2\pi^2}\Gamma,\nonumber\\
 \frac{\partial^2 \mathcal{G}_{x}}{\partial \Lambda_y^2}\bigg|_{1^+}-\frac{\partial^2 \mathcal{G}_{x}}{\partial
 \Lambda_y^2}\bigg|_{1^-}\!&\!\simeq\!&\!
\frac{\sqrt{3} }{2\pi^2}\Gamma,
\end{eqnarray}
where $\Gamma$ is given in Eq.~(\ref{A10}).

From Eq.~(\ref{correlation002}), we can obtain
\begin{eqnarray}
\mathcal{G}_{y}\big|_{1^+}\!&\!=\!&\!\mathcal{G}_{y}\big|_{1^-},\nonumber\\
\frac{\partial \mathcal{G}_{y}}{\partial
\Lambda_x}\bigg|_{1^+}\!&\!=\!&\!\frac{\partial
\mathcal{G}_{y}}{\partial
 \Lambda_x}\bigg|_{1^-},\nonumber\\
 \frac{\partial \mathcal{G}_{y}}{\partial \Lambda_y}\bigg|_{1^+}\!&\!=\!&\!\frac{\partial \mathcal{G}_{y}}{\partial
 \Lambda_y}\bigg|_{1^-},\nonumber\\
 \frac{\partial^2 \mathcal{G}_{y}}{\partial \Lambda_x^2}\bigg|_{1^+}-\frac{\partial^2 \mathcal{G}_{y}}{\partial
 \Lambda_x^2}\bigg|_{1^-}\!&\!\simeq\!&\!
-\frac{\sqrt{3} }{2\pi^2}\Gamma,\nonumber\\
 \frac{\partial^2 \mathcal{G}_{y}}{\partial \Lambda_x\partial \Lambda_y}\bigg|_{1^+}
-\frac{\partial^2 \mathcal{G}_{y}}{\partial \Lambda_x\partial
\Lambda_y}\bigg|_{1^-}\!&\!\simeq\!&\!
\frac{\sqrt{3} }{2\pi^2}\Gamma,\nonumber\\
 \frac{\partial^2 \mathcal{G}_{y}}{\partial \Lambda_y^2}\bigg|_{1^+}-\frac{\partial^2 \mathcal{G}_{y}}{\partial
 \Lambda_y^2}\bigg|_{1^-}\!&\!\simeq\!&\!
-\frac{\sqrt{3} }{2\pi^2}\Gamma.
\end{eqnarray}

\end{document}